\documentclass[a4paper,11pt]{article}
\usepackage{pos}
\usepackage{wrapfig}
\usepackage{physics}

\title{Critical behavior towards the chiral limit at
vanishing and non-vanishing chemical potentials}
\ShortTitle{Critical behavior towards the chiral limit}

\author[a]{Olaf Kaczmarek}
\author[a]{Frithjof Karsch}
\author[a]{Anirban Lahiri}
\author[b]{Sheng-Tai Li}
\author*[a,c]{Mugdha Sarkar}
\author[a]{Christian Schmidt}
\author[d]{Philipp Scior}

\affiliation[a]{Fakult\"at f\"{u}r Physik, Universit\"{a}t Bielefeld,
 Bielefeld, Germany}
\affiliation[b]{ Key Laboratory of Quark \& Lepton Physics (MOE) and Institute of
Particle Physics, Central China Normal University, Wuhan 430079, China}
\affiliation[c]{Department of Physics, National Taiwan University, Taipei 10617, Taiwan} 
\affiliation[d]{Physics Department, Brookhaven National Laboratory, Upton, NY 11973, USA}

\emailAdd{okacz@physik.uni-bielefeld.de}
\emailAdd{karsch@physik.uni-bielefeld.de}
\emailAdd{alahiri@physik.uni-bielefeld.de}
\emailAdd{lishengtai@mails.ccnu.edu.cn}
\emailAdd{mugdha@physik.uni-bielefeld.de}
\emailAdd{schmidt@physik.uni-bielefeld.de}
\emailAdd{pscior@bnl.gov}

\abstract{We study the scaling behavior of the (2+1)-flavor QCD crossover region towards the chiral limit with smaller-than-physical light quark mass gauge ensembles, generated using the HISQ fermion discretization. At zero chemical potential, we study the fluctuations of conserved charges and their correlations with the chiral condensate,
towards the chiral limit. We analyse the role of universal and regular contributions to the above quantities. We find a preliminary estimate of the leading curvature coefficient of the chiral phase transition line using scaling arguments.}

\FullConference{%
 The 38th International Symposium on Lattice Field Theory, LATTICE2021
  26th-30th July, 2021
  Zoom/Gather@Massachusetts Institute of Technology
}

\begin{document}
\maketitle

\section{Introduction}
Knowledge of the nature of the chiral phase transition is a 
crucial piece of the puzzle of understanding the QCD phase diagram. 
With advances in high performance computing, lattice simulations 
with dynamical light quarks having smaller-than-physical 
masses have become feasible. This allows 
us to hunt for the signals of criticality towards 
the chiral limit, if any. A recent review of the motivations 
and the current developments of this field can be found in 
Ref. \cite{Lahiri:2021lrk}.

\begin{wrapfigure}{r}{0.4\textwidth}
  \begin{center}
    \includegraphics[width=.35\textwidth]{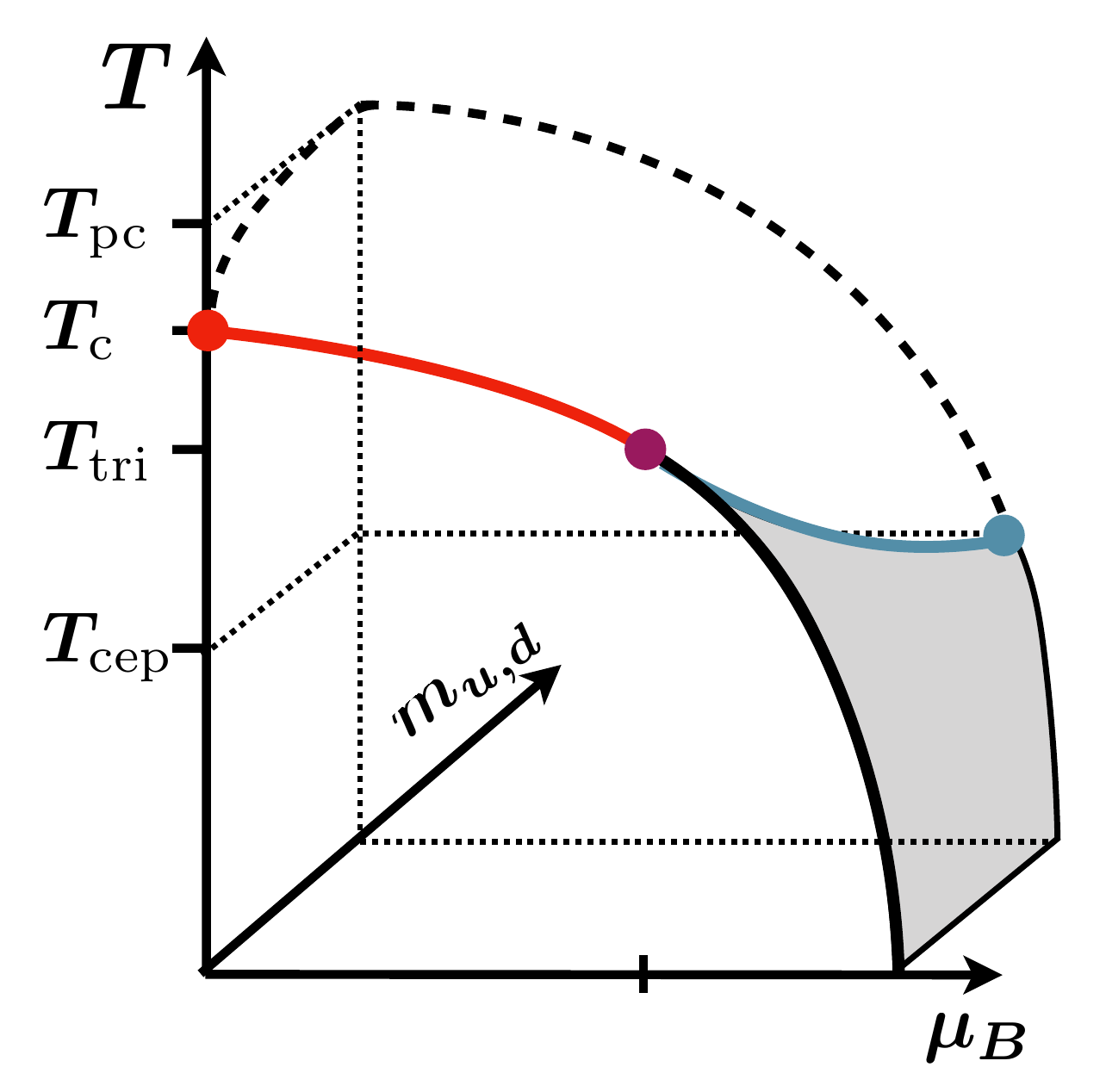}
  \end{center}
  \caption{Schematic phase diagram for (2+1)-flavor QCD \cite{Karsch:2019mbv} in the temperature ($T$), baryon chemical potential ($\mu_B$) and light quark mass ($m_l$) directions.}
  \label{3dphasediag}
\end{wrapfigure}
In the schematic 3d QCD phase diagram depicted in 
Fig. \ref{3dphasediag} (taken from Ref.~\cite{Karsch:2019mbv}), 
we are interested in the chiral 
limit of the two degenerate light quarks 
($m_{u,d}\equiv m_l =0$) at vanishing chemical potential, which 
is accessible to lattice calculations. The chiral phase 
transition temperature $T_c$ (denoted by the red dot in figure) 
was recently determined to be around 
132 MeV \cite{HotQCD:2019xnw}, which is below 
the crossover temperature $T_{pc}$ at physical 
quark masses \cite{HotQCD:2018pds,Borsanyi:2020fev}. 
However, the order of the chiral phase transition is still 
not clear beyond doubt. Depending on 
the restoration of the anomalously broken $U_A(1)$ 
symmetry around $T_c$, one possibility is that 
the chiral phase transition is a second order 
transition belonging to the 3d O(4) universality class 
\cite{Pisarski:1983ms}, which seems to be preferred 
from recent studies 
\cite{HotQCD:2019xnw, Clarke:2020htu, Bazavov:2019www, Ding:2020xlj, Cuteri:2021ikv} 
over other possibilities.

In this proceedings, we present our investigations of the
conserved charge fluctuations calculated with smaller-than-physical
light quark masses. The imprint of the criticality on these 
observables is important from the viewpoint of the 
heavy-ion collision experiments \cite{Ejiri:2005wq,Friman:2011pf}.
We further report on mixed observables which are derivatives 
of the free energy density with respect to the light quark mass 
and chemical potential. Next, we present a preliminary calculation 
of the curvature of the chiral phase transition line at 
non-vanishing chemical potentials.
The curvature in the chiral limit is important for locating 
the supposed tricritical point at $T=T_\mathrm{tri}$, 
denoted by the maroon dot at the end 
of the red second order transition line in Fig. \ref{3dphasediag}. 
The tricritical point in turn affects the location of 
the critical endpoint $T_\mathrm{cep}$ \cite{Hatta:2002sj} 
(denoted by the blue point in Fig. \ref{3dphasediag}), 
which is being actively searched for in experiments.

\section{Lattice setup}
The gauge ensembles have been generated with HISQ 
fermion discretization and tree-level
improved Symanzik gauge action. The
ensembles were generated starting from 
physical light quark mass $m_l = m_s/27$ to 
smaller-than-physical light quark masses
$m_l$ = $m_s/40$, $m_s/80$, $m_s /160$, keeping 
strange quark mass $m_s$ fixed at physical values, 
with corresponding pion masses of 
140 MeV, 110 MeV, 80 MeV and 58 MeV.
For scale setting, we use the kaon decay
constant obtained in calculations with
the HISQ action, i.e., $f_K$ = $156.1/\sqrt{2}$ MeV 
\cite{Bazavov:2010hj}. We present results from 
measurements done at the largest simulated volumes 
for each mass at fixed time extent $N_\tau$ = 8
with aspect ratios $N_\sigma/N_\tau$ in the range $4-7$.

\section{Critical behavior of thermodynamic quantities}
In Wilsonian RG theory, the couplings of the Hamiltonian 
near a fixed point in the infinite coupling space can be 
classified into those that respect the symmetry 
(which gets broken across the critical point) and those that 
break the symmetry explicitly. With respect to the chiral 
phase transition in hot and dense (2+1)-flavor QCD, 
the temperature $T$, chemical potentials $\mu_X$ for 
conserved charges $(X=B,Q,S)$, etc.\footnote{The strange quark 
mass $m_s$ may as well be considered to be an energy-like 
coupling as we shall do later.}, define the 
symmetry-preserving \textit{energy-like} scaling field $t$ 
and the light quark mass $m_l$ defines the symmetry-breaking 
\textit{magnetic-like} scaling field 
$h$. In the vicinity of the critical point, 
the dimensionless scaling fields are defined as 
\begin{equation}{\label{Eq:t}}
    t = \frac{1}{t_0}\left(\frac{T-T_c}{T_c} + \kappa_2^X\left(\frac{\mu_X}{T}\right)^2\right), \qquad
    h \equiv \frac{H}{h_0} = \frac{1}{h_0}\frac{m_l}{m_s},
\end{equation}
where $t_0, h_0$ and $\kappa_2^X$ are dimensionless non-universal 
constants and $\kappa_2^X$ also denotes the leading order 
curvature of the chiral phase transition line. 

Using the linearized approximation of the scaling fields 
near the critical point, 
we may express the free energy 
density $f$ as the sum of singular (non-analytical) and 
regular contributions as \cite{Friman:2011pf},
\begin{equation}{\label{Eq:free}}
    \frac{f(t,h)}{T^4} = Ah^{(2-\alpha)/\beta\delta}f_f(z) + \mathrm{regular~terms},
\end{equation}
where $f_f(z)$ is a universal scaling function of the 
scaling variable $z\equiv t/h^{1/\beta\delta}$, 
$\alpha, \beta$ and $\delta$ are the critical 
exponents of the universality class and $A$ is a 
non-universal constant. With higher derivatives of $f$, 
the singular part becomes dominant and diverges in the 
chiral limit. The various scaling functions have been 
studied precisely for 3d O(2) \cite{Engels:2000xw} and 
O(4) \cite{Engels:2011km} universality classes and have 
been successfully used to fit lattice data 
\cite{Clarke:2020htu,Ejiri:2009ac,Kaczmarek:2011zz}. 
This indicates the possibility that the chiral phase 
transition belongs to the O(4) universality class. 
Since we work at finite lattice spacing with staggered
quarks, we use 3d O(2) critical exponents in this study.
Given the fact that the lattices used in this project 
are large enough \cite{HotQCD:2019xnw}, we use the 
infinite volume scaling functions in our analyses for any 
thermodynamic observable.

\subsection{Fluctuations of conserved charges}\label{sec:fluct}

\begin{figure}
\centering
\includegraphics[width=.45\textwidth]{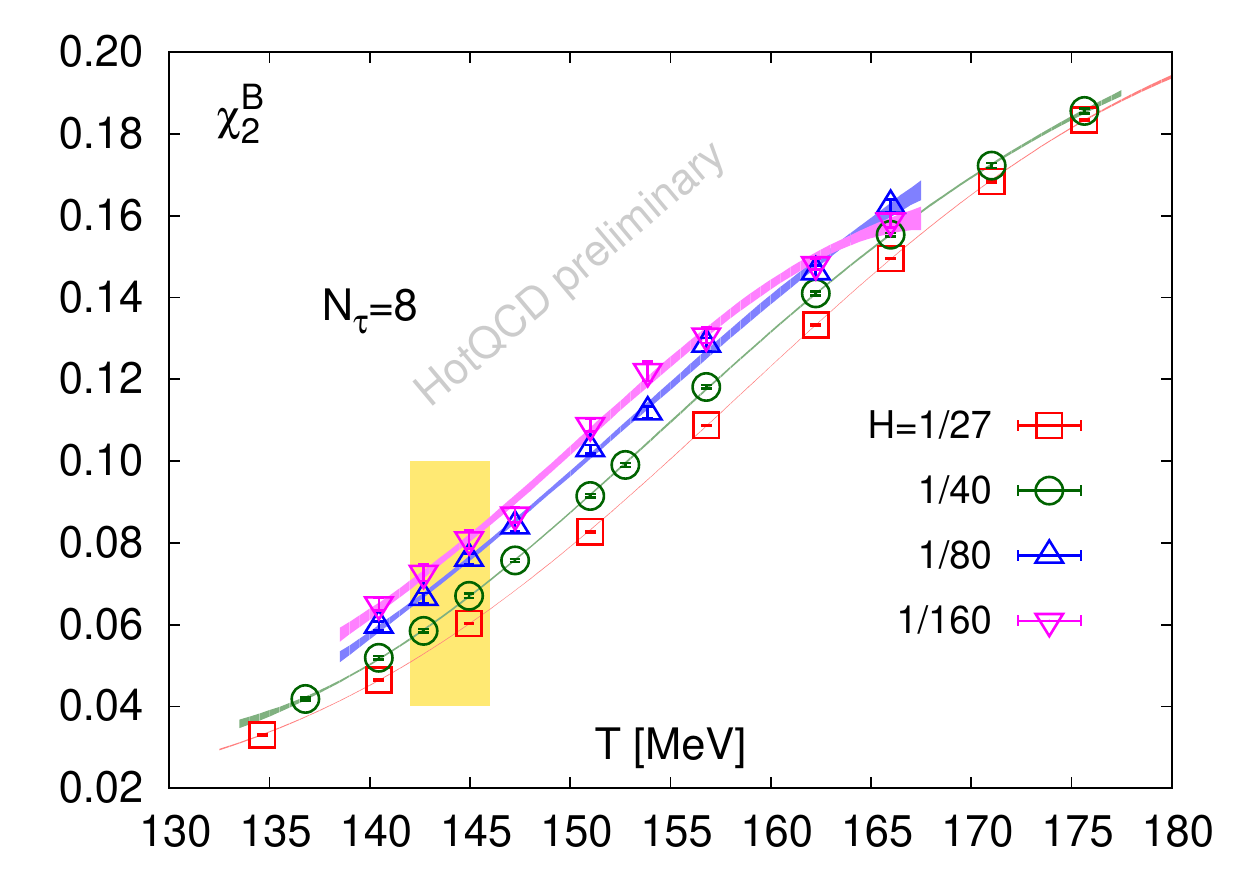}
\includegraphics[width=.45\textwidth]{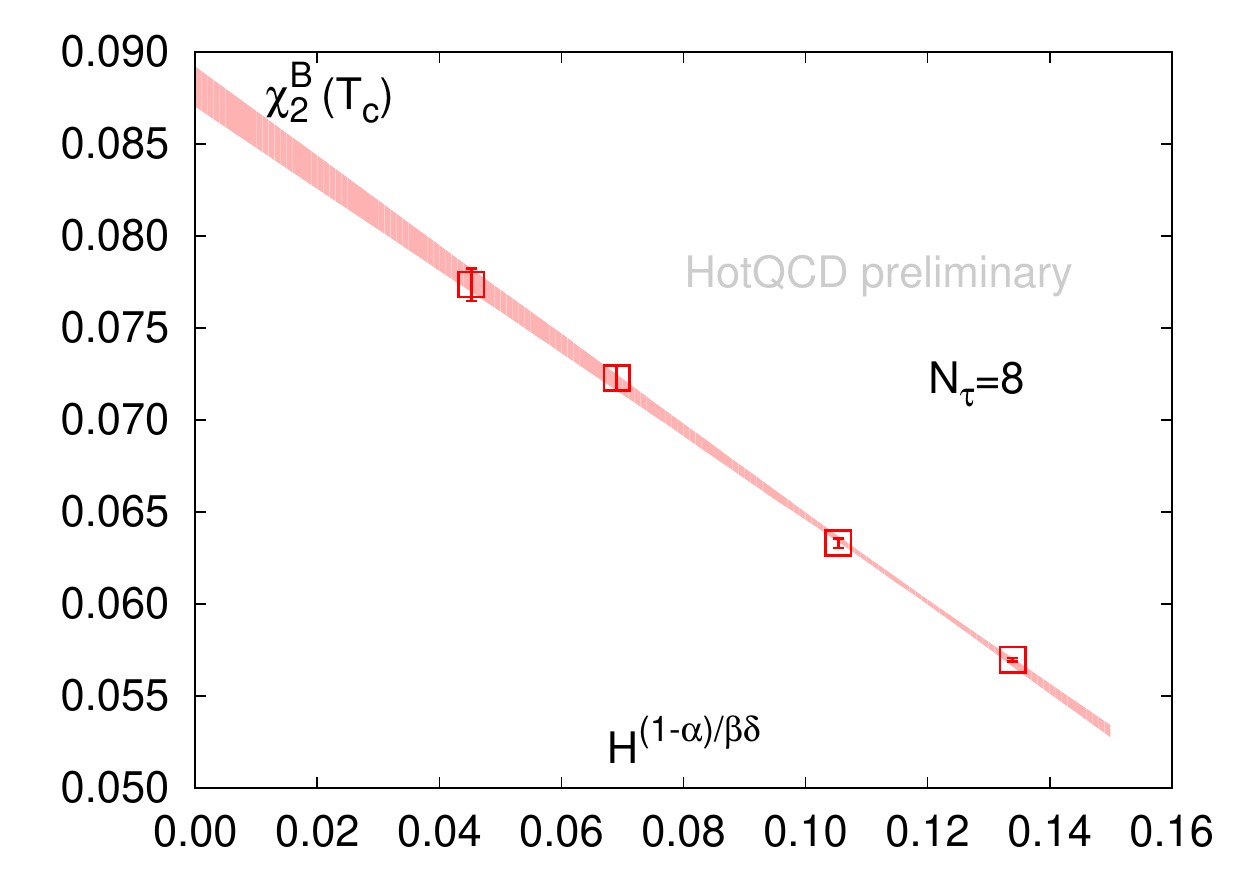}
\caption{Left : The second order cumulant of baryon 
number fluctuations, $\chi_2^B$, as a function of the 
temperature at various quark mass ratios 
$H = m_l/m_s$. Right : The values of $\chi_2^B$ at 
$T=T_c^{N_\tau=8}$ plotted against 
$H^{(1-\alpha)/\beta\delta} = H^{0.61}$ (using 3d O(2) 
critical exponents).}
\label{fig:chi2B}
\end{figure}

\begin{figure}
\centering
\includegraphics[width=.45\textwidth]{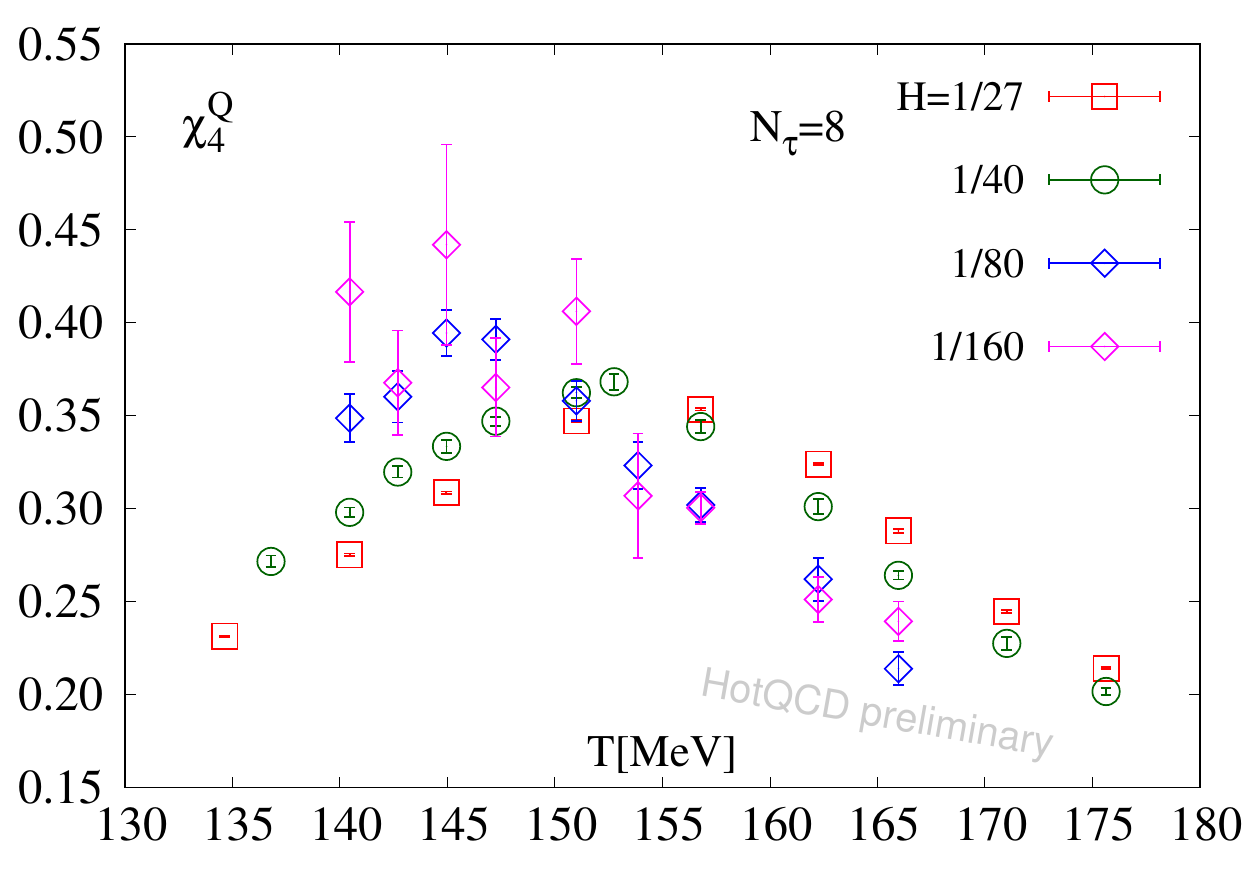}
\includegraphics[width=.45\textwidth]{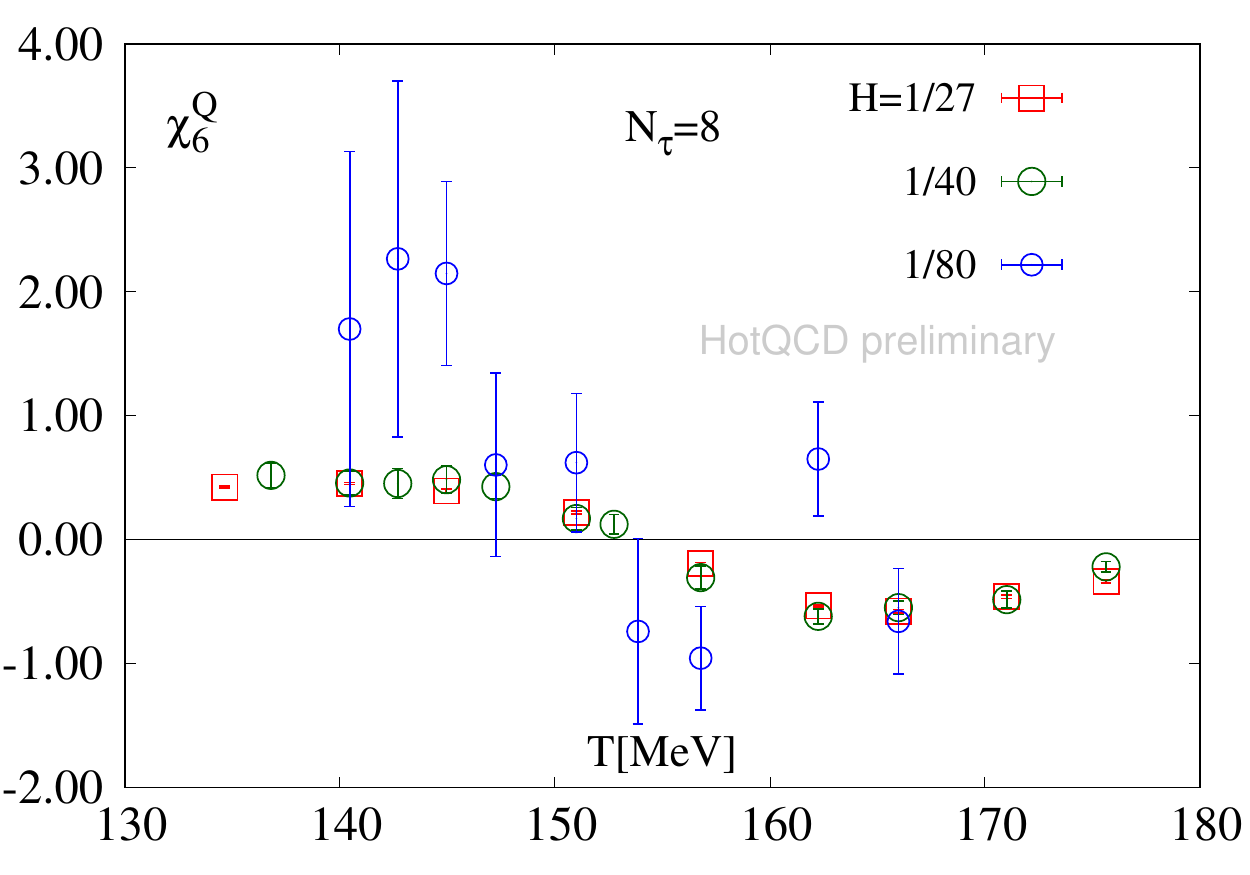}
\caption{Left : The $4^{\rm th}$ order cumulant of 
electric charge fluctuations, $\chi_4^Q$, as a function of the 
temperature for different $H$ values. Right : The same as left 
but for the $6^{\rm th}$ order cumulant, $\chi_6^Q$. The data for 
$H=1/160$ is still too noisy.}
\label{fig:chi4Q}
\end{figure}

We are interested in the behavior of conserved charge 
fluctuations at vanishing chemical potential as 
we move towards the chiral limit. From Eq. \ref{Eq:free},
the scaling behavior of the cumulants can be written as
\begin{equation}{\label{Eq:fluct}}
    \chi_{2n}^X \equiv -\frac{\partial^{2n} f/T^4}{\partial(\mu_X/T)^{2n}}
    \Bigg|_{\mu_X=0} = - A_n(2\kappa_2^X)^n h^{(2-\alpha-n)/\beta\delta} f_f^{(n)}(z)
    + \mathrm{regular~terms},
\end{equation}
where $f_f^{(n)}(z)$ are derivatives of the universal 
free energy scaling function with respect to (w.r.t) 
$z$ and $A_n$ are non-universal constants.
From Eq. \ref{Eq:t}, it is easy to show that 
two derivatives of $f$ w.r.t
$\mu_X$ yields the same singular part as a single 
derivative w.r.t $T$ up to a constant. Hence, 
the $(2n)^{\rm th}$ order cumulants are actually 
$n^{\rm th}$ order derivatives of $f$ w.r.t $T$ in 
terms of scaling 
behavior. For 3d O(N) models in general, $\alpha$ is 
negative and the divergence starts from $6^{\rm th}$ 
order onwards. The second order cumulants behave as 
the energy density and the fourth order cumulants 
behave as the specific heat which should develop a 
characteristic spike around $T_c$ in the chiral limit 
\cite{Cucchieri:2002hu}. 
We reported on this behavior in a previous proceeding 
\cite{Sarkar:2020soa}. Here we present the updated 
statistics in Figs \ref{fig:chi2B} and \ref{fig:chi4Q}.

The singular part of the second order cumulants at $T=T_c$ 
for different $H$ could be estimated from a scaling fit which 
is linear in $H^{(1-\alpha)/\beta\delta}$, as can be seen from 
Eq.~\ref{Eq:fluct} and from the right-hand plot in Fig.~\ref{fig:chi2B}. 
The ratios among the singular parts of different second order 
cumulants can be used to estimate the ratio of the curvatures 
along corresponding $\mu_X$ directions \cite{Sarkar:2020soa}.

The features of the $4^{\rm th}$ and $6^{\rm th}$ order 
fluctuations are governed by
the scaling functions $f''_f(z)$ and $f'''_f(z)$, respectively. 
These scaling functions for the 3d O(4) universality class can 
be found in Ref. \cite{Friman:2011pf}. We present updated results for 
the $4^{\rm th}$ and $6^{\rm th}$ order cumulants of 
the electric charge fluctuations in Fig. \ref{fig:chi4Q}, 
where the improvement of the former one is vivid and the apparent
increase in the peak height, according to the scaling expectation, 
can be better realized. 
Compared to Ref. \cite{Sarkar:2020soa}, we have been able to add 
$H=1/80$ data to the $\chi_6^Q$ calculation but it is evident that 
we still require more statistics.

\subsection{Mixed observables}
The mixed derivatives of the free energy density 
{\it i.e.}~derivatives w.r.t both energy-like 
($t$) and magnetic-like ($h$) couplings are divergent 
already from second order onward. 
We study two classes of 
mixed observables, corresponding to $m_s$ derivatives of 
the light quark chiral condensate 
$\Sigma_l$, {\it i.e.} $m_s \partial \Sigma_l/\partial m_s$, 
and $\mu_X$ derivatives of $\Sigma_l$, {\it i.e.} $C^{\Sigma_l}_{2,X}\equiv
{\partial^2 \Sigma_l}/{\partial \left(\mu_X/T\right)^2}$,
respectively.
Since the strange mass $m_s$ does not break the 2-flavor chiral 
symmetry, we consider it to be a energy-like coupling.
The dimensionless light quark chiral condensate is defined in 
the $f_K$ scale as 
\begin{equation}{}
    \Sigma_l = -\frac{m_s}{f_K^4}\frac{\partial f}{\partial m_l} \; .
\end{equation} 
Since we do not take the continuum limit in this work, 
we need not worry about the ultraviolet divergence in $\Sigma_l$.
In terms of scaling behavior, $\Sigma_l$ can be expressed as
\begin{equation}{}
    \Sigma_l = h^{1/\delta} f_G(z) + \mathrm{regular~terms}\,,
\end{equation}
where $f_G(z)$ is a universal scaling function 
related to $f_f(z)$ as $f_G(z)=-(1+1/\delta)f_f(z) + 
(z/\beta\delta)f'_f(z)$. 
Upon taking two derivatives w.r.t. $\mu_X$, 
we have the following scaling expectation,
\begin{equation}{\label{Eq:C2Bsigma}}
    C^{\Sigma_l}_{2,X} = 2\kappa_2^X h^{(\beta-1)/\beta\delta} f'_G(z) 
    + \mathrm{regular~terms}\,,
\end{equation}
where the singular term is divergent for both $O(2)$ 
and $O(4)$ critical exponents. The mixed observable 
$m_s \partial \Sigma_l/\partial m_s$ has the same singular 
behavior up to a constant but with different regular 
contributions. It may be noted that $C^{\Sigma_l}_{2,X}$
is the leading order coefficient in the Taylor expansion 
of the light quark chiral condensate $\Sigma_l$ in 
chemical potential $\mu_X$. We refer to \cite{HotQCD:2018pds,Steinbrecher:2018jbv} 
for the techniques used in the computation of $\Sigma_l$ 
and its Taylor expansion coefficients.

\begin{figure}
\centering
\includegraphics[width=.45\textwidth]{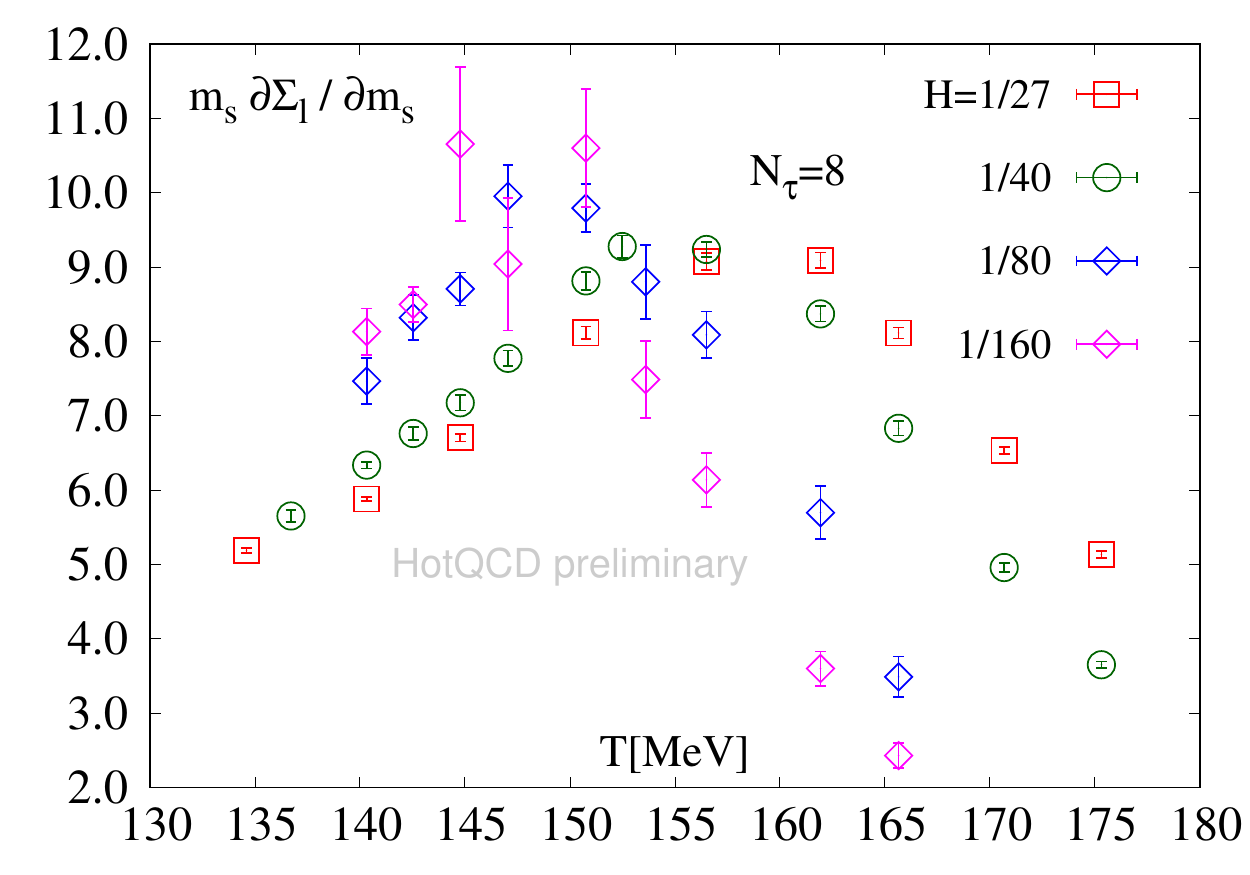}
\includegraphics[width=.45\textwidth]{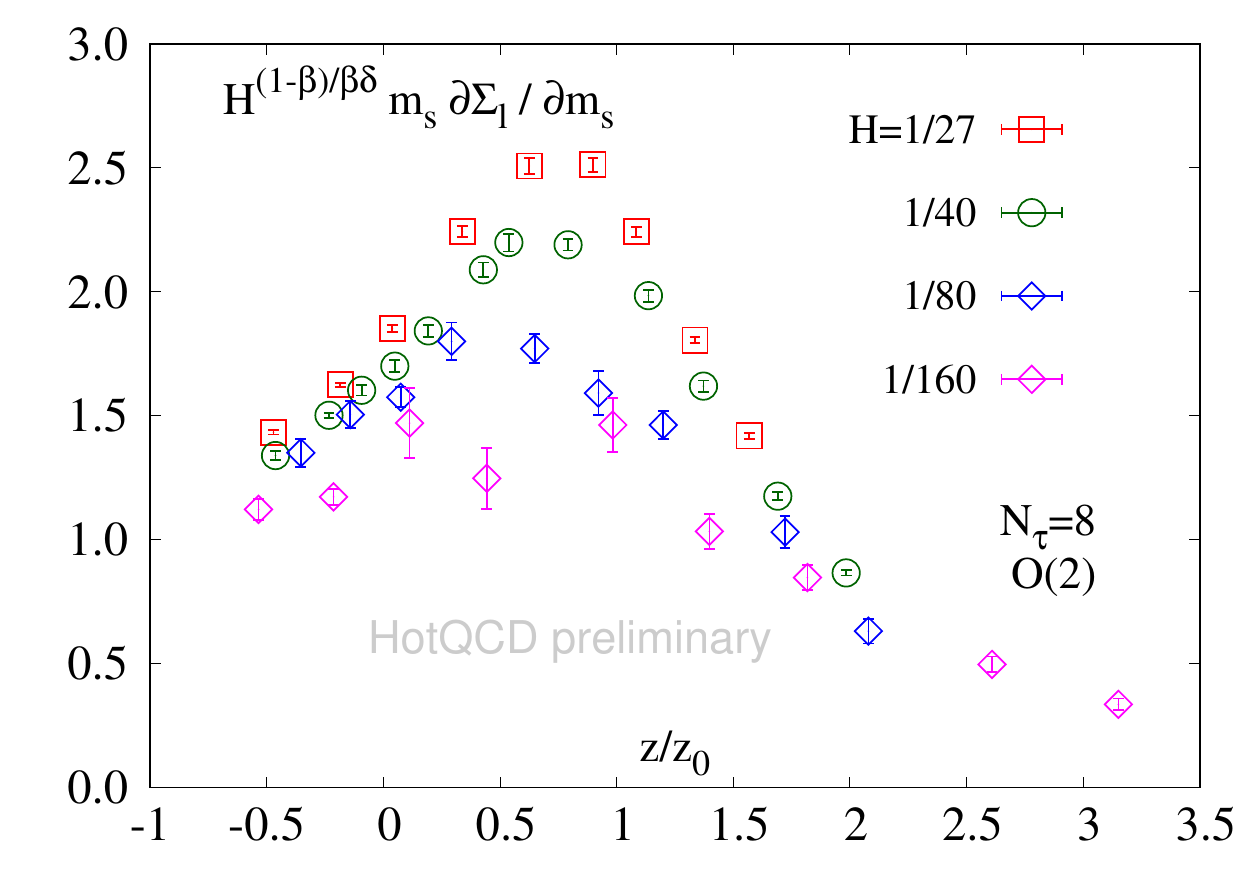}
\caption{Left : The dimensionless strange mass $m_s$ 
derivative of the light quark condensate $\Sigma_l$,
$m_s \partial \Sigma_l/\partial m_s$, \textit{versus} 
temperature for various $H$ values. Right : The same 
quantity scaled by the factor $H^{(1-\beta)/\beta\delta}$
and plotted against $z/z_0 = H^{-1/\beta\delta}(T-T_c)/T_c$ 
for different $H$ values.}
\label{fig:chils}
\end{figure}
\begin{figure}
\centering
\includegraphics[width=.45\textwidth]{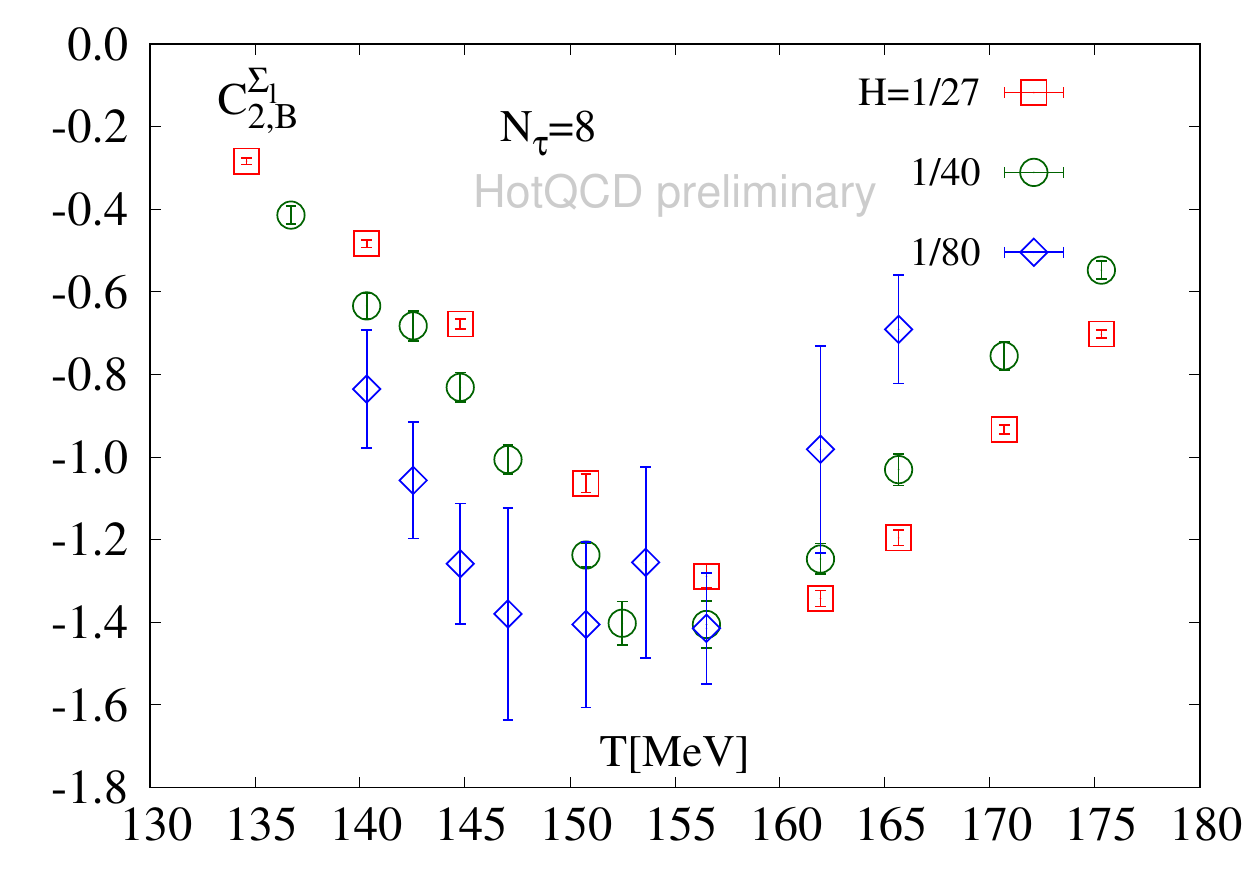}
\includegraphics[width=.45\textwidth]{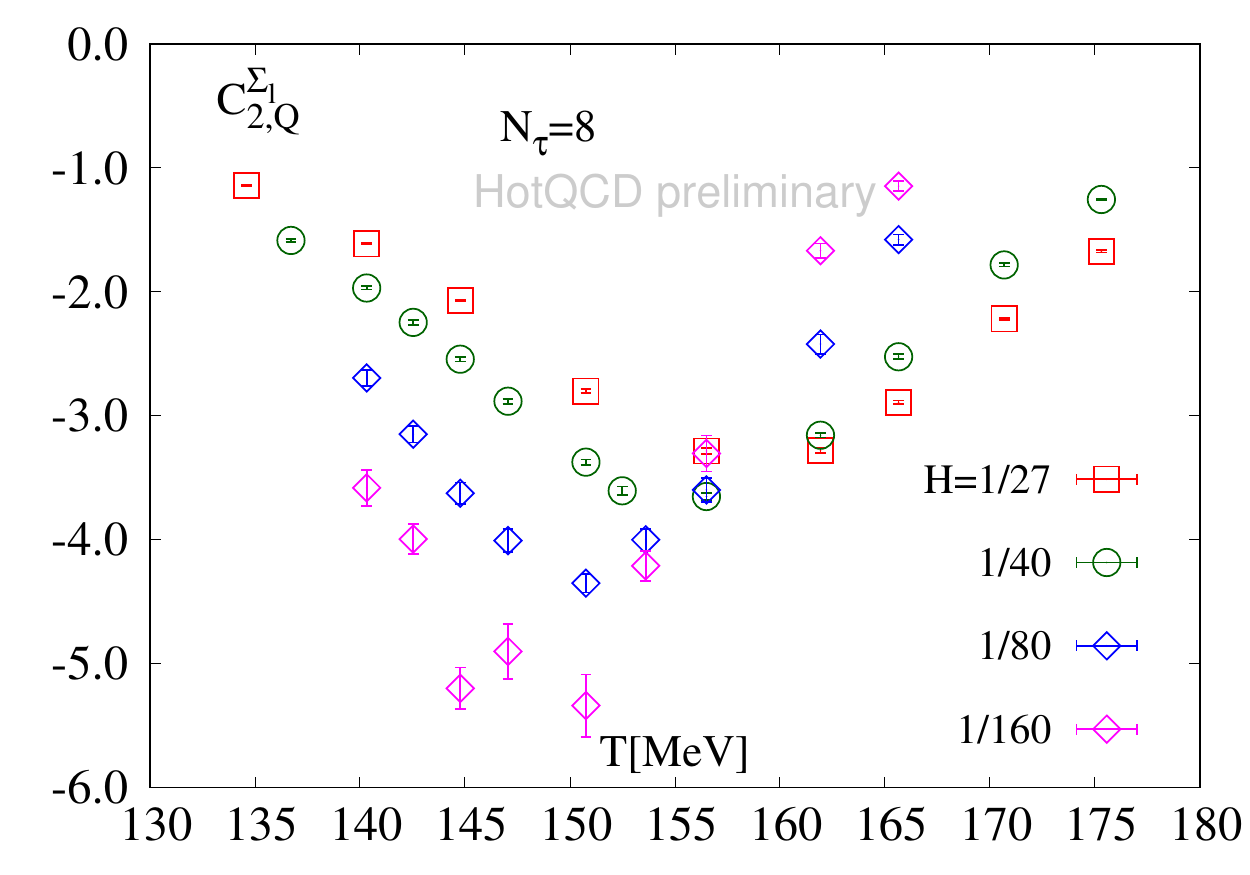}
\includegraphics[width=.45\textwidth]{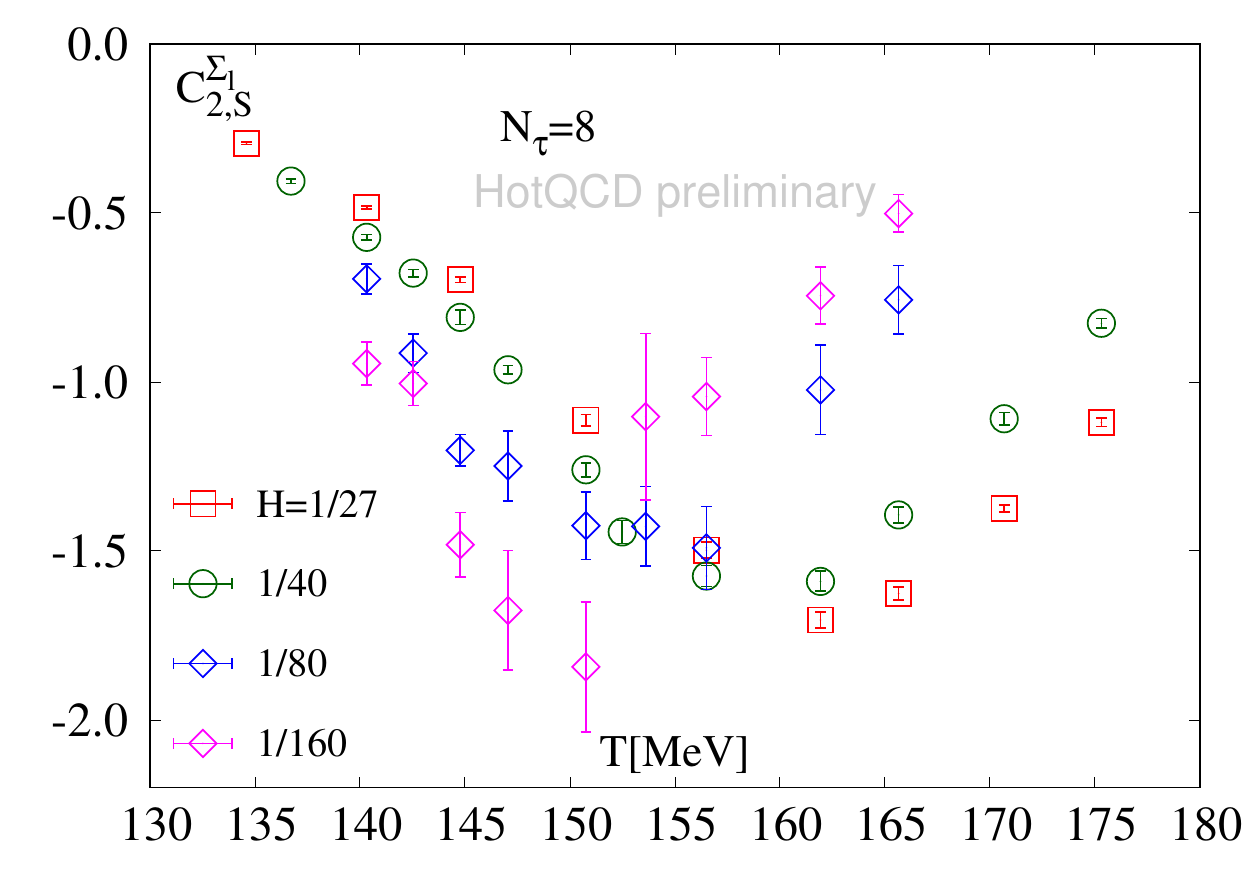}
\caption{Top left : The second order Taylor coefficient, 
$C^{\Sigma_l}_{2,B}$, in the expansion of $\Sigma_l$ around 
baryon number chemical potential $\mu_B=0$, plotted against 
temperature at given $H$ values.
Top right : The same coefficient of the expansion in electric 
charge chemical potential $\mu_Q$. Bottom : The same coefficient 
for strangeness chemical potential $\mu_S$.}
\label{C2BQS_pbp_plots}
\end{figure}

We present our results for $m_s \partial \Sigma_l/\partial m_s$ 
and $C^{\Sigma_l}_{2,X}$ in the left panel of 
Fig.\ \ref{fig:chils} and in Fig.\ \ref{C2BQS_pbp_plots}, 
respectively. In the right plot of 
Fig. \ref{fig:chils}, we show 
$m_s \partial \Sigma_l/\partial m_s$ scaled by 
$H^{(1-\beta)/\beta\delta}$ plotted against $z/z_0$. It 
can be easily seen from Eq. \ref{Eq:C2Bsigma} that the 
scaled plot represents the scaling function $f'_G(z)$ 
up to a constant, in the absence of regular 
contributions. If we are close to the chiral limit, 
the regular terms would be negligible in comparison 
to the divergent singular part and we should observe 
the scaled data as a function of $z/z_0$ to fall on 
top of each other for different 
$H\to 0$. However, we do not observe such a behavior 
for our current $H$ values in Fig. \ref{fig:chils}, 
which indicates that the regular contributions are 
not negligible. As discussed in Ref. \cite{Clarke:2020htu}, 
very similar observables w.r.t
scaling expectation may behave quite differently in the lattice data,
due to regular terms. We are currently investigating the role 
of various regular terms in our data. It should also be 
noted that we had presented results for some mixed 
observables in the $T$ normalization previously 
\cite{Sarkar:2020soa}, as opposed to $f_K$ normalization 
in this proceeding. The change in 
normalization produces a qualitative change in the scaling 
behavior and we intend to discuss on this in more detail in 
a future study.

\section{Curvature of the chiral phase transition line}
From the definition of the reduced temperature $t$ in 
Eq. \ref{Eq:t}, one can easily see that the condition 
$t(T,\mu_X) = 0$ maps out the chiral phase transition 
line in the $T-\mu_X$ plane. Using this condition,
we can write the chiral transition temperature as a 
function of the chemical potential as
\begin{equation}
    T_c (\mu_X) = T_c(0) \left(1 - \kappa_2^X \left(\frac{\mu_X}{T}\right)^2\right) 
    \; . 
\end{equation}
It is clear from the above that the coefficient 
$\kappa_2^X$ determines the leading-order curvature 
of the chiral transition line for small $\mu_X$ 
values.

One can try to find an estimate of $\kappa_2^X$ in the 
following way. For any observable $\mathcal{O}$ at $T=T_c$ and 
$\mu_X=0$, we can rewrite the 
chemical potential and temperature derivatives, 
using Eq. \ref{Eq:t}, as,
\begin{align}{}
    \frac{\partial^2 \mathcal{O}}{\partial \left(\mu_X/T\right)^2}\Bigg|_{(T_c,0)} &= 
    \frac{2\kappa_2^X}{t_0} \frac{\partial \mathcal{O}}{\partial t} \; , \label{Eq:dfdmu}\\
    \frac{\partial \mathcal{O}}{\partial T}\Bigg|_{(T_c,0)} &= \frac{1}{t_0 T_c}
    \frac{\partial \mathcal{O}}{\partial t} \; . \label{Eq:dfdT}
\end{align}
Combining Eqs. \ref{Eq:dfdmu} and \ref{Eq:dfdT}, 
we find an estimate for the chiral curvature at small 
values of chemical potential,
\begin{equation}{\label{Eq:k2}}
    \kappa_2^X = \frac{T^2}{2T_c}\frac{\left({\partial^2 \mathcal{O}}/{\partial \mu_X^2}\right)\Big|_{(T_c,0)}}{\left({\partial \mathcal{O}}/{\partial T}\right)\Big|_{(T_c,0)}} \; .
\end{equation}
At finite values of $H$, the derivatives of $\mathcal{O}$ may 
consist of regular contributions along with the singular 
part, which add corrections to $\kappa_2^X$.
We choose $\mathcal{O}$ such that the singular parts diverge 
towards the chiral limit to obtain better estimates 
for $\kappa_2^X$ from our data at small $H$ 
values. One such choice is the chiral order parameter. 
We consider the light quark chiral condensate $\Sigma_l$ 
in this discussion\footnote{We could have chosen instead the subtracted 
condensate $\Sigma$, but the presence of the strange 
quark condensate term may alter the singular behavior 
leading to non-trivial corrections in $\kappa_2^X$.}. 
The scaling expectation of the $\mu_X$ derivative,  $C^{\Sigma_l}_{2,X}\equiv{\partial^2 \Sigma_l}/{\partial \left(\mu_X/T\right)^2}$, is given in Eq. \ref{Eq:C2Bsigma}.
As discussed in Sec. \ref{sec:fluct}, the $T$ 
derivative of $\Sigma_l$ should be the same as 
Eq. \ref{Eq:C2Bsigma} except for factors of $\kappa_2^X$ and 
different regular terms. 

\begin{figure}
\centering
\includegraphics[width=.45\textwidth]{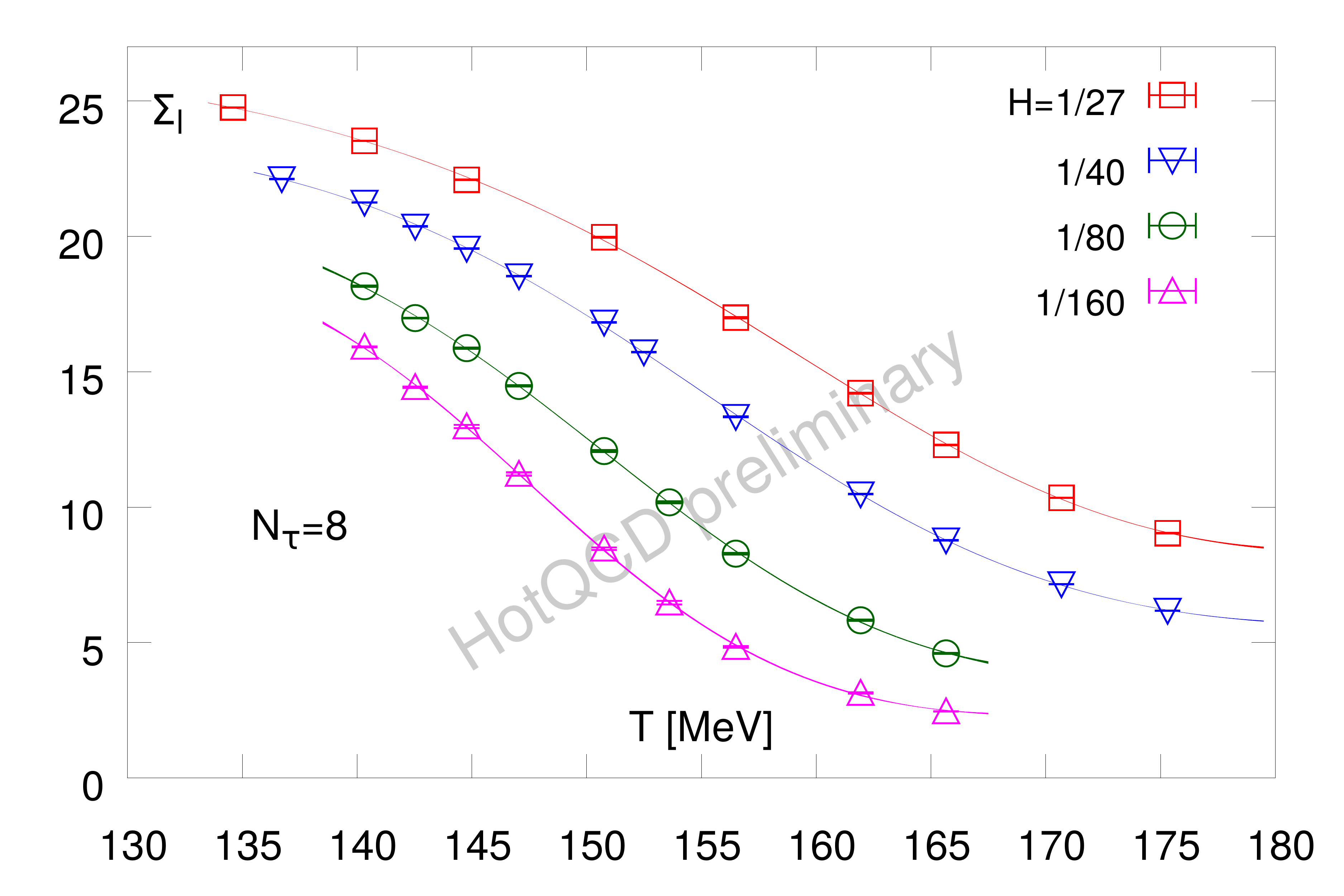}
\includegraphics[width=.45\textwidth]{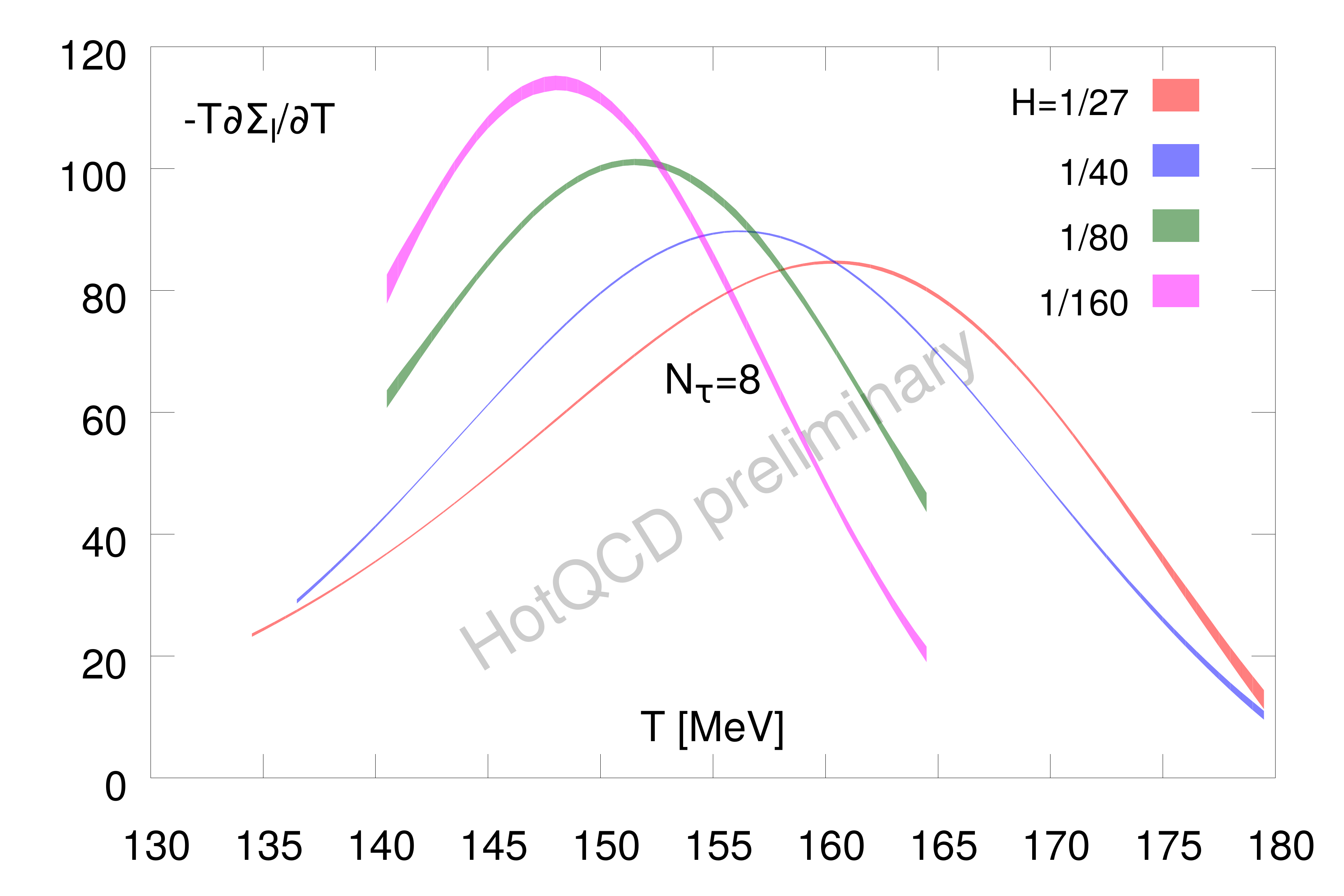}
\caption{Left : The light quark chiral condensate $\Sigma_l$ 
\textit{versus} temperature, obtained at different $H$ values. 
The data as a function of temperature has been interpolated 
using rational polynomials.
Right : The temperature derivative, 
$-T\partial\Sigma_l/\partial T$, obtained from 
the fit of $\Sigma_l$ shown in the left panel, is plotted 
against the temperature for various $H$ values.}
\label{ddT_M_Mb_plots}
\end{figure}
\begin{figure}
\centering
\includegraphics[width=.45\textwidth]{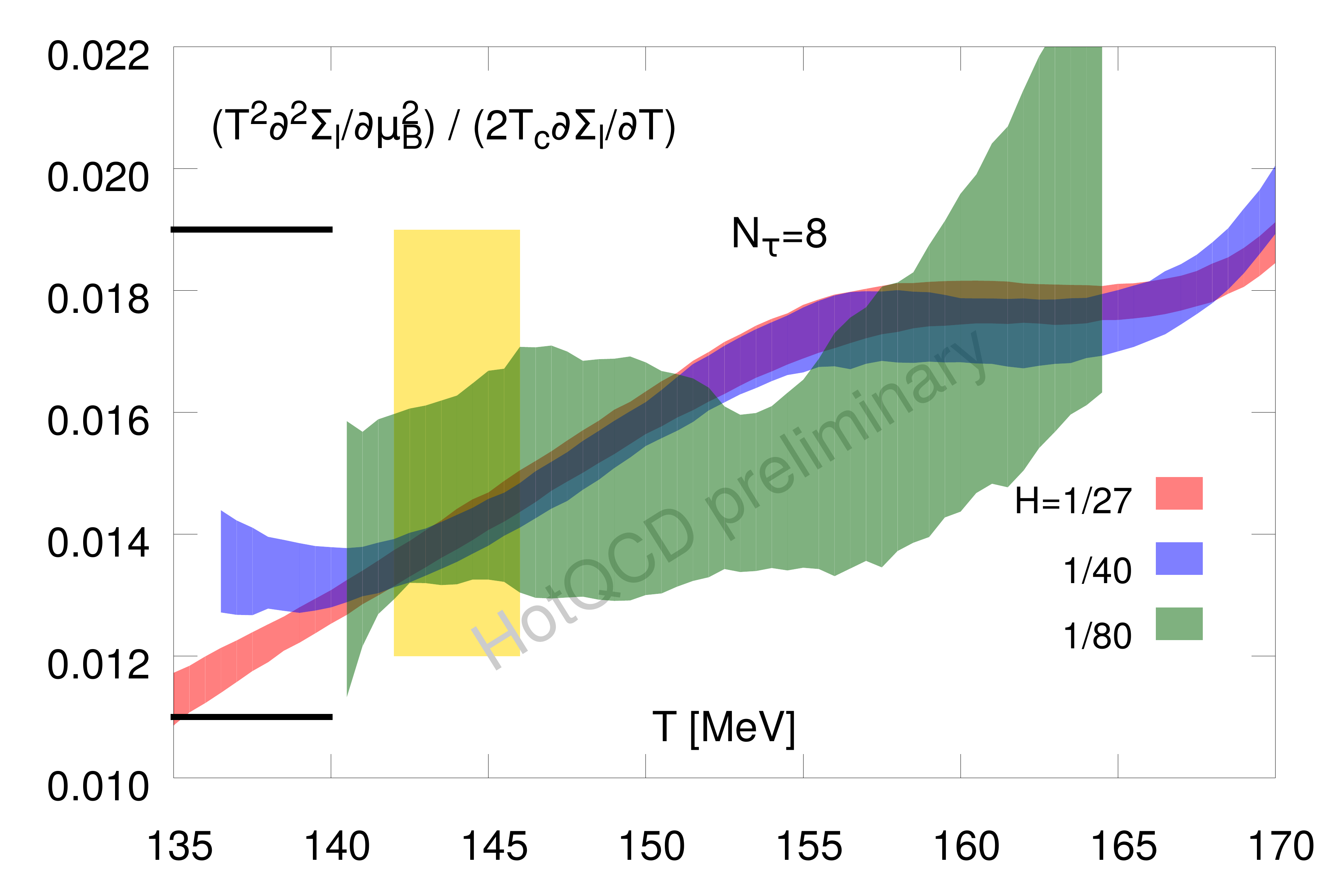}
\includegraphics[width=.45\textwidth]{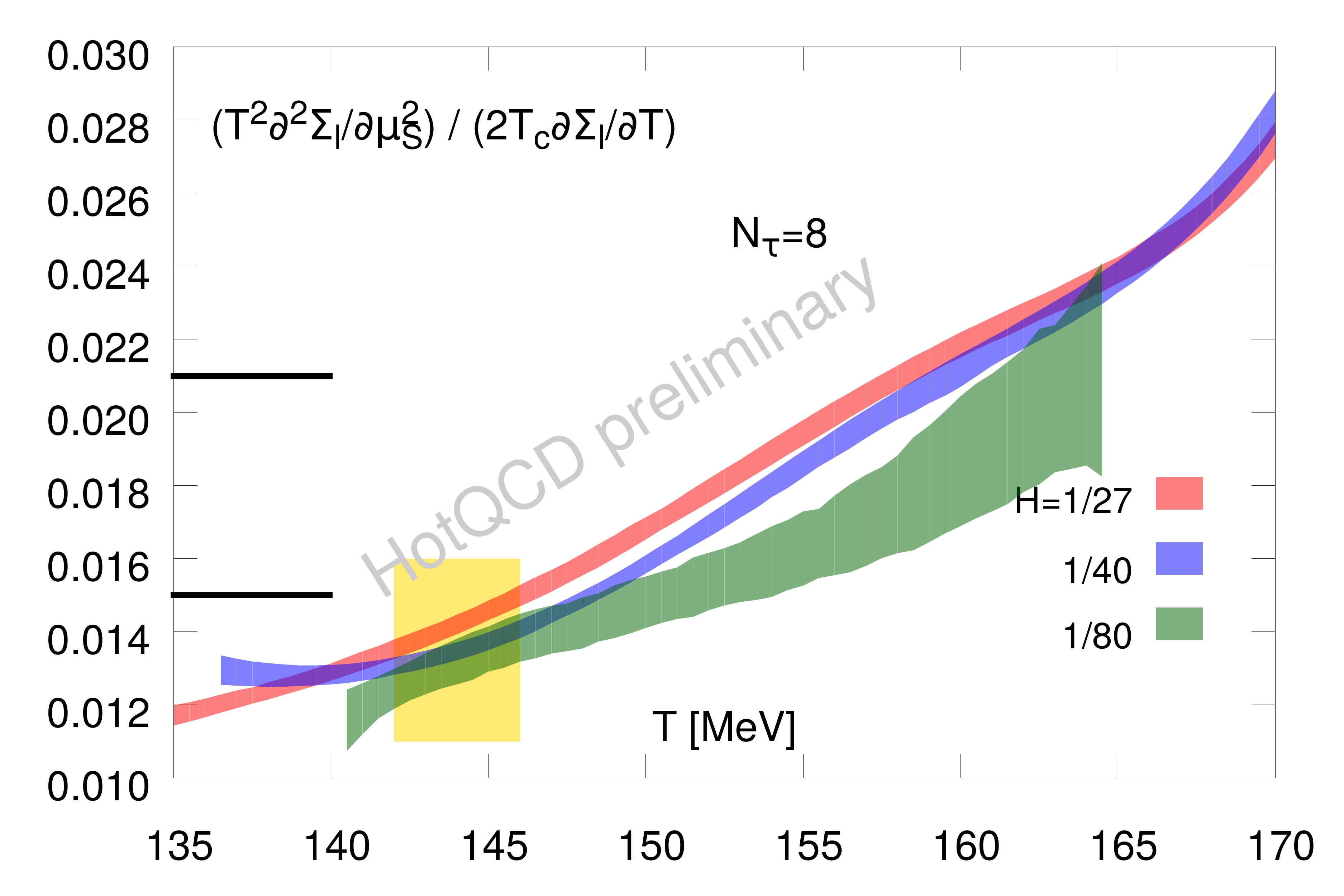}
\caption{The ratio of the second order chemical potential derivative 
$\partial^2 \Sigma_l/\partial \mu_X^2$ and temperature derivative 
$\partial \Sigma_l/\partial T$ of the light quark condensate 
$\Sigma_l$ across the temperature range for various $\mu_X, X=B,S$. 
The chiral limit estimates of $\kappa_2^X$ get extracted in 
the yellow band around $T_c$. For comparison, 
we denote the 68\% confidence interval of the 
continuum extrapolated results of the 
leading order curvature at physical mass, determined at 
$T_{pc}$ for different $\mu_X$ \cite{HotQCD:2018pds}, 
by black bars along the $y$-axis.}
\label{k2BQS_Mb_plots}
\end{figure}

We already discussed our lattice results for 
$C^{\Sigma_l}_{2,X}$ at smaller-than-physical 
light quark masses in Fig. \ref{C2BQS_pbp_plots} in 
the previous section. 
To calculate the $T$ derivative of $\Sigma_l$, 
we first interpolate $\Sigma_l$
using rational polynomials, as shown in the 
left panel of Fig.\ \ref{ddT_M_Mb_plots},
and then took the $T$ derivative of the interpolating 
function to get $\partial\Sigma_l/\partial T$
which is shown in the right panel of Fig.\ \ref{ddT_M_Mb_plots}.
Finally, we also interpolate the data of the $\mu_X$ derivatives 
and take the ratio of Figs.\ \ref{C2BQS_pbp_plots} 
and Fig.\ \ref{ddT_M_Mb_plots} to obtain the estimates of 
the curvature values shown in Fig. \ref{k2BQS_Mb_plots}. 
The interpolations as well as the ratio calculation have been
done on fake samples generated under Gaussian approximation
with mean and standard deviation being the average and the
uncertainty, respectively, of a particular data point.
We read off the chiral curvature $\kappa_2^X$ from 
the temperature interval around the chiral phase 
transition temperature on lattices with temporal 
extent $N_\tau =8$, $T=T_c^{N_\tau =8} \simeq144$ MeV
(determined in Ref.\ \cite{HotQCD:2019xnw}).
The continuum extrapolated results for the curvature
of the crossover line along various chemical potentials 
at physical masses \cite{HotQCD:2018pds}
has been indicated by black bars in Fig.~\ref{k2BQS_Mb_plots}
(see figure caption).
Our preliminary results suggest that the chiral limit 
result for the curvature coefficient  $\kappa_2^B$ 
remains almost unchanged from the physical mass curvature 
which may be argued from the relatively small change in 
the nucleon masses towards the chiral limit
\cite{Young:2009zb}. 
Similar calculation for $\mu_S$ is shown in the right 
panel of Fig.\ \ref{k2BQS_Mb_plots} and the corresponding chiral
curvature coefficient $\kappa_2^S$ at $N_\tau=8$ appears 
to be below the corresponding continuum physical mass value.
The curvature coefficient of the chiral transition line 
along various chemical potentials can also be obtained 
as a fit parameter from the scaling fit of the $\mu_X$ 
derivative of the chiral condensate with the ansatz 
given in Eq. \ref{Eq:C2Bsigma}. This has been tried in a 
previous work using the p4-action of staggered
quarks \cite{Kaczmarek:2011zz} and we plan to repeat 
the same analysis with our current HISQ data. 
This will give a more complete picture about the mass 
dependence of the (pseudo-)critical lines towards the 
chiral limit.

The same ratio in Eq. \ref{Eq:k2}, except for the 
factor $T^2/2T_c$ replaced by $T_{pc}/2$, was computed 
for thermodynamic observables like the pressure and energy 
density at physical mass $H=1/27$ in Ref. 
\cite{Bazavov:2017dus}. The ratio represents the curvature 
of the line of constant physics (LCP) for the observables 
computed at $T=T_{pc}$ and $\mu_B=0$. It was found that they 
agree quite well with the curvature of the crossover line 
along $\mu_B$. From Fig. \ref{k2BQS_Mb_plots},
we can also extract the curvature of 
the LCP of the chiral order parameter
for $H=1/27$ at $T=T_{pc}^{N_\tau=8} \simeq 161$ MeV
and that seems to be in well agreement with previous determination
of the curvature of the pseudo-critical line using various
pseudo-critical conditions.

\section{Conclusions and outlook}
Our results at finite lattice spacing appear 
consistent with a chiral phase transition 
belonging to the 3d, O(2) universality class 
(3d, O(4) in the continuum). The scaling behavior of 
second order conserved charge fluctuations is similar 
to that of an energy density which allows us to separate 
the singular and regular contributions of these quantities 
at physical or any given masses. The ratios of the singular 
contributions provide an estimate of the ratio of the 
chiral limit curvatures, and it seems to indicate that 
the curvatures along various chemical potential 
directions depend weakly on the light quark mass. 
We expect the strange quark mass is 
an energy-like coupling w.r.t the 2-flavor chiral phase 
transition and the strange quark condensate to behave like 
an energy density. However, at our current $H$ values,
we need more detailed analyses including the regular 
contributions to understand the scaling behavior of 
the mixed observable, $m_s \partial \Sigma_l/\partial m_s$ 
better.

We provided a first preliminary result for the estimators 
of the curvature of the phase transition line 
in the chiral 
limit with the HISQ action. In agreement with the ratios 
of the curvatures from the second order cumulants, 
we find that the curvature changes only weakly as we move 
towards the chiral limit.

Next, we plan to do a scaling analysis 
with input from the previously known 3d, O(2) scaling 
functions to understand the interplay of the singular 
and regular contributions in the observables. In addition, 
the fluctuations at smaller-than-physical masses in the 
hadronic phase shall 
be compared to predictions from the HRG model, which 
by itself cannot capture the critical behavior. In the 
future, we plan to do an extensive study at different 
lattice spacings and volumes for different masses to 
take the chiral, continuum and thermodynamic limits.

\section*{Acknowledgements}

This work was supported by the Deutsche Forschungsgemeinschaft (DFG,
German Research Foundation) Proj.~No.~315477589-TRR 211; and by the
German Bundesministerium f\"ur Bildung und Forschung through Grant No.~05P18PBCA1. 
We thank the HotQCD for providing access to their latest data
sets and for many rewarding discussions.

\bibliographystyle{JHEP}
\bibliography{main}

\end{document}